\documentclass{elsart}
\usepackage{graphicx}
\usepackage[dvips]{epsfig}
\usepackage{dcolumn}
\usepackage{bm}
\usepackage{amssymb}
\usepackage{amsmath}

\begin{document}
\begin{frontmatter}

\title{On one-dimensional models for hydrodynamics}
\date{}

\author{Carlos Escudero\corauthref{cor}}
\corauth[cor]{Corresponding author, phone: +44 (0) 1865 283891, fax: +44 (0) 1865 273583.}
\ead{escudero@maths.ox.ac.uk}

\address{
Mathematical Institute, University of Oxford, 24-29 St Giles', Oxford OX1 3LB, United Kingdom}

\begin{abstract}
To date it has not been possible to prove whether or not the three-dimensional incompressible Euler equations develop singular behaviour in finite time. Some possible singular scenarios, as for instance shock-waves, are very important from a physical point of view, since they imply the connection among the macroscopic and the microscopic scale. Therefore, the appearence of this type of singularity or a similar one might be interpreted as a possible explanation of the transition to turbulence. In order to
clarify the question, some one-dimensional models for ideal incompressible hydrodynamics have been introduced and analyzed,
and it was proven that shock-waves appear in finite time within this type of flow. In this work we question the validity
of these models and analyze the physical meaning that the occurrence of a singularity in an incompressible flow, if it
happens, may have.
\end{abstract}

\end{frontmatter}

\section{Introduction}

It is not yet known whether or not the three-dimensional incompressible Euler equations develop singular behaviour in finite time. Far from being a pure mathematical problem, it has deep connections with some fundamental questions in physics.
Actually, the relation between the appearance of finite time singularities and the onset of turbulence has been conjectured, because singularity formation may be a mechanism of energy transfer from large to small scales
(or may not, see~\cite{onsager,frisch}). There are several types of blow-ups that can occurr in a fluid flow, such as, for instance, an infinite increment of the velocity in some spatial point. This kind of blow-up is not useful for an understanding of turbulence, since it might be interpreted as a break down of the nonrelativistic description. This is, if the velocity becomes too large, it is necessary to modify the equations to include relativistic effects~\cite{landau}. On the other hand, a shock-wave is a singularity of the first spatial derivative of the velocity. Shock-waves are events in the flow where the continuum description breaks down and a kinetic description, such as the Boltzmann equation, becomes necessary~\cite{cercignani,caflish}. This suggests that shock-waves are a link between the microscopic and the macroscopic scale. In conclusion, we may argue that the onset of turbulence might be related with such situations in which the flow develops a shock-wave (or a similar phenomenon, see below), but in which the velocity remains bounded. All along this work we will be concerned (unless explicitely indicated) with flows carrying a finite total kinetic energy, since we consider this characteristic as necessary to keep the physical meaning of the flow.

First of all, let us clarify a question of language. Usually, it is understood that a shock-wave is a discontinuity of the velocity of the flow. Of course, the first spatial derivative of the flow will present a Dirac delta singularity located at the shock, but it is possible to find a singularity of the first spatial derivative of a continuous velocity field, just by letting the slope of the tangent go to infinity. We can name this type of events (continuous velocity and divergence of its first spatial derivative) as quasi-shock-waves, due to its high similarity to traditional shock-waves. Let us explain why.
Consider a continuous one-dimensional velocity field $v(x)$, its first spatial derivative at $x_0$ is given by
$v'(x_0)=\lim_{h \to 0}(v(x_0+h)-v(x_0))/h$. This quantity will be divergent if there is a high accumulation of different values of $v$ near $x_0$. A high accumulation of different values of the velocity in the neighbourhood of one point reveals that the continuum description of the fluid is less accurate, and this descripcion becomes ill defined when a divergence is developed. This shows that shock-waves and quasi-shock-waves are singular events of the very same kind, both revealing the failure of the continuum description of the fluid flow and the need of a kinetic description (and maybe the connection between the macroscopic and the microscopic scales). An equivalent way to describe the phenomenon is to see it as the collision of two fluid particles, as we will see below.

\section{First approach to singular behaviour}

The three-dimensional incompressible Euler equations read:
\begin{subequations}
\begin{equation}
\label{euler}
\partial_t v + (v \cdot \nabla)v = - \nabla p,
\end{equation}
\begin{equation}
\label{incompresibility}
\nabla \cdot v = 0,
\end{equation}
\end{subequations}
where $v$ is the velocity, a three-dimensional vector, and $p$ is the pressure, a scalar. Eq.(\ref{euler}) is Newton's law
for the fluid, the left hand side is the convective derivative of the velocity and the right hand side is the force in terms of the pressure. Eq.(\ref{incompresibility}) is the incompressibility condition for the fluid. These equations can be obtained from the Boltzmann equation performing the Chapman-Enskog expansion~\cite{chapman}, but full mathematical proofs are available only before the appearance of singularities~\cite{bardos,raymond}. The Boltzmann equation describes the dynamics of a rarefied gas, taking into account two basic processes: the free flight of the particles and
their collisions. In the hydrodynamic description, the convective derivative simulates the free flight of the particles and the gradient of the pressure is the first order correction to the interaction among them. The second order correction yields the Navier-Stokes equations, that are obtained by adding the term $\mu \nabla^2 v$ to the right hand side of Eq.(\ref{euler}), where $\mu>0$ is the viscosity of the fluid. The third order is still an open problem, however some
insight has been gained into it by means of a regularization of the Chapman-Enskog expansion~\cite{rosenau,escudero}.
These partial achivements suggest that at small scales the viscosity loses effectivity~\cite{escudero}, what makes more interesting to study the dynamics of an inviscid fluid when we are close to the onset of turbulence.

Due to the mathematical difficulties that a direct treatment of the Euler equations imply, many simplified models have been introduced in order to understand better Euler dynamics. One classical approximation is the Burgers equation:
\begin{equation}
\partial_t v + v \partial_x v = 0,
\end{equation}
where $v$ is the one-dimensional velocity. It is interesting to note that Burgers equation is a one-dimensional analogue of
Euler equations where the pressure has been supressed.
This equation is known to develop shock-waves in finite time as a consequence of the crossing of its characteristics~\cite{evans}.
This result can be easily understood if one remembers the form of the Hamilton-Jacobi equation for the free particle
\begin{equation}
\label{hamilton}
\partial_t S = \frac{1}{2m} (\partial_x S)^2,
\end{equation}
recalling that the momentum is given by $p=\partial_x S$ we see that deriving once with respect to the spatial variable Eq.(\ref{hamilton}), substituting $p=mv$, and reversing the direction of time $t \to -t$ (remember that classical mechanics is invariant with respect to time reversing)
we recover Burgers equation. Thus Burgers equation can be thought as the evolution of the velocity of a cloud of free particles. Since this gas is noninteracting, the characteristics of the Hamilton-Jacobi equation, the Hamilton equations (the particle trajectories), will cross when subject to an adequate initial condition, what physically means that the
fluid particles collide in finite time.
Obviously, this fact does not require an infinite kinetic energy or an unbounded velocity of the fluid.
This cannot be directly translated to the Euler equations, since the pressure term acts as a force against this type of collisions. However, if we introduce an interaction between the particles in the form of a viscosity
\begin{equation}
\partial_t v + v \partial_x v = \mu \nabla^2 v,
\end{equation}
we get global existence of the solution. So it seems that in this case the shocks are an artifact of the noninteracting character of the particles.

It is important to understand what we mean by fluid particles and by collisions among them. A fluid particle is not
a molecule composing the fluid. It is a coarse-graining concept that corresponds to a point in the hydrodynamical
(macroscopic) description of the fluid. But it is actually composed of a high number of molecules. Correspondingly,
a collision among fluid particles does not implie the collapse of two molecules. It is only the superposition of
two coarse-graining entities reflecting the failure of the macroscopic approach. It is then necessary to describe
the fluid microscopically in terms of the actual molecules composing the flow (this is the kinetic decription we refer
to above).

It is known that the compressible Euler equation develop shock-waves as well~\cite{sideris} (and that they are again regularized by the viscosity in one dimension). This shocks are, however, completely due to the compressible character of the flow, and cannot be extrapolated to incompressible flows~\cite{sideris}. This suggests that, if a singularity is present in an incompressible flow, it would not be of the shock type. The real problem is in contrast much more complex: when small-scale structures appear through the nonlinear dynamical evolution, they tend to display, at least locally, a much faster dependence on one particular spatial dimension (a phenomenon known as depletion of nonlinearity)~\cite{frisch2}. In the limit, a flow collapsed into one spatial dimension displays a singular behaviour of the quasi-shock-wave type: the fluid particles collapse into a one-dimensional structure revealing the failure of the continuum description. Also, there is rigorous work showing that if a singularity is present, it is due to the collapse of a small-scale structure~\cite{constantin5}. This shows that if a divergence appears in the three-dimensional Euler flow, it might be a phenomenon related to a shock-wave, although not necessarily a shock-wave itself.

To clarify the question at hand, let us consider one example of infinite energy exact solution that shows a strong tendence towards the shock-wave. Some exact infinite energy singular solutions might be found in~\cite{gibbon}, and among them we choose the following one in cylindrical coordinates
\begin{equation}
v=v(v_r(r,t),0,v_z(z,t)),
\end{equation}
without a swirling component of the velocity, and where
\begin{subequations}
\begin{equation}
v_r(r,t)=\frac{1}{2}\frac{r}{T^*-t},
\end{equation}
\begin{equation}
v_z(z,t)=-\frac{z}{T^*-t},
\end{equation}
\end{subequations}
where $T^*$ is the blow-up time. For a better understanding of the fluid dynamics one can integrate the equations of motion for the fluid particles
\begin{subequations}
\begin{equation}
\frac{dr}{dt}=v_r(r,t),
\end{equation}
\begin{equation}
\frac{dz}{dt}=v_z(z,t),
\end{equation}
\end{subequations}
that yield
\begin{subequations}
\begin{equation}
r(t)=r(0)\sqrt{\frac{T^*}{T^*-t}},
\end{equation}
\begin{equation}
\theta(t)=\theta(0),
\end{equation}
\begin{equation}
z(t)=z(0)\frac{T^*-t}{T^*}.
\end{equation}
\end{subequations}
This clearly indicates that all the particles of the fluid collapse in the plane $z=0$ at the finite time $t=T^*$. In spite of this strong tendence to form a shock, this type of blow-up cannot be considered as a shock-wave since the
collapse occurs in a spatial point at an infinite distance from the origin, and at an infinite velocity. What this
solution indicates is that the nonrelativistic description of the flow has broken down.

\section{Model equations for the vorticity dynamics}

One of the most important results proved about the regularity of the solutions of the Euler equations is the Beale-Kato-Majda theorem~\cite{beale}, that says that the solution exists globally in time if and only if
\begin{equation}
\int_0^T \max_x |\omega(x,s)|ds < \infty,
\end{equation}
where $\omega=\nabla \times v$ is the vorticity. This made very interesting to study the evolution of the vorticity in the Euler equations:
\begin{equation}
\partial_t \omega + (v \cdot \nabla) \omega = \omega \cdot \nabla v,
\end{equation}
where the velocity can be recovered from the Biot-Savart Law:
\begin{equation}
v(x)=\frac{1}{4 \pi} \int_{\mathbb{R}^3} \frac{x-y}{|x-y|^3}\times \omega(y) dy.
\end{equation}
This equation can also be expressed in the following way
\begin{equation}
\frac{D \omega}{Dt}=\mathcal{D}(\omega)\omega,
\end{equation}
where $D\omega/Dt=\partial_t \omega + (v \cdot \nabla)\omega$ is the convective derivative of the vorticity, and $\mathcal{D}$ is a symmetric matrix given by
\begin{equation}
\mathcal{D}=(\mathcal{D}_{ij})=\left[\frac{1}{2}\left(\frac{\partial v_i}{\partial x_j}+\frac{\partial v_j}
{\partial x_i}\right)\right].
\end{equation}
The operator relating $\omega$ to $\mathcal{D}\omega$ is a linear singular integral operator that commutes
with translation. In one spatial dimension there is only one such operator: the Hilbert transform.

In this spirit, Constantin {\it et al.}~\cite{constantin3} proposed the following one-dimensional model for the vorticity equation:
\begin{equation}
\partial_t \omega = H(\omega) \omega,
\end{equation}
where $H(\omega)$ is the Hilbert transform of $\omega$:
\begin{equation}
H(\omega) = \frac{1}{\pi} \mathrm{P.V.} \int_{-\infty}^\infty \frac{\omega(y)}{x-y}dy,
\end{equation}
and P.V. denotes the Cauchy principal value integral
\begin{equation}
\mathrm{P.V.} \int_{-\infty}^\infty f(x) dx = \lim_{\epsilon \to 0} \int_{|x| \ge \epsilon} f(x) dx.
\end{equation}
This equation was solved explicitly and it was shown that it blows up for some finite time $T_0$.
However, it has been proven that the viscous analogue of this equation
\begin{equation}
\partial_t \omega = H(\omega) \omega + \epsilon \omega_{xx},
\end{equation}
blows up for some finite time $T_\epsilon$ such that $T_\epsilon < T_0$; this is, adding diffusion makes the solution less regular. This is unsatisfactory
in view of the result by Constantin~\cite{constantin4}, which says that if the solution to the Euler equations is smooth then the solutions to the
slightly viscous Navier-Stokes equations are also smooth. In order to prevent this behaviour, De Gregorio~\cite{gregorio1,gregorio2} introduced an improved
model keeping the convective derivative:
\begin{equation}
\partial_t \omega + u \partial_x \omega = \omega H(\omega) + \mu \partial_{xx} \omega
\end{equation}
with viscosity $\mu \ge 0$. This equation does not develop singular behaviour, and De Gregorio concluded that one-dimensional
models for hydrodynamics are not able to faithfully represent three-dimensional incompressible flow.

Baker, Li, and Morlet studied a very similar one-dimensional model simulating vortex sheet dynamics
~\cite{baker}:
\begin{equation}
\label{ccf}
\partial_t \theta  - H(\theta) \partial_x \theta = \mu \theta_{xx},
\end{equation}
which has been reinterpreted as a model for the Euler equations by C\'ordoba, C\'ordoba, and
Fontelos~\cite{acordoba} after switching $\mu=0$. In this model $\theta$ is a scalar carried by the flow,
and the vorticity and the velocity are defined, respectively, by $\omega=\theta_x$ and $v=-H(\theta)$.
In reference~\cite{acordoba} it is proven that this system blows up for some finite time in some spatial
point, provided the initial condition is even,
positive and compactly supported; then $\lim_{t \to T^*}||\theta_x||_{L^\infty}=\infty$ for some
$T^*<\infty$.
As proven there, the solution is even whenever the initial condition is so, implying that its Hilbert transform, say the velocity, is odd, and thus null at the origin. Also, the transport character of this equation implies that $||\theta||_{L^\infty}(t)=||\theta||_{L^\infty}(0)$.
Numerical evidence has indicated that the blow-up appears in the origin~\cite{acordoba,morlet}, where a
cusp of $\theta$ is formed. This type of singularity corresponds to a (quasi-)shock, but we still need to know
if the velocity is bounded at the origin to assure that a genuine (quasi-)shock-wave is formed. It is an important fact that the positivity of $\theta$ implies that $v=-H(\theta)$ is a decreasing function, and since it is antisymmetric, this flow simulates the collision of two fluid jets coming from infinite with opposite directions. The collision point is the
origin, and the (quasi-)shock is generated when two fluid particles collide there.

We can establish the boundedness of the velocity using the properties of the space of functions of
bounded mean oscillation (BMO) (for the basic properties of this space see, for instance, \cite{stein}):
\begin{equation}
||v||_{\mathrm{BMO}} \le C ||\theta||_{L^\infty},
\end{equation}
for a finite positive constant $C$, and
\begin{equation}
||v||_{\mathrm{BMO}} = \sup_{Q} \frac{1}{|Q|} \int_Q |v-v_Q|dx,
\end{equation}
where $Q$ is any closed interval of $\mathbb{R}$, and $v_Q = \int_Q v dx$.
From the very definition of the BMO norm we see that for any $\epsilon>0$, the inequality
\begin{equation}
||v||_{\mathrm{BMO}} \ge \frac{1}{2\epsilon} \int_{-\epsilon}^{\epsilon} |v-v_\epsilon| dx 
\end{equation}
holds, and now we can use the fact that $v$ is an odd function, and thus $v_\epsilon=\int_{-\epsilon}^\epsilon v dx=0$, leading us to claim that
\begin{equation}
||v||_{\mathrm{BMO}} \ge \frac{1}{2\epsilon} \int_{-\epsilon}^{\epsilon} |v| dx.
\end{equation}
This implies the estimate
\begin{equation}
\frac{1}{2\epsilon} \int_{-\epsilon}^{\epsilon} |v| dx \le C ||\theta||_{L^\infty},
\end{equation}
homogeneous in $\epsilon$, so we can take the limit $\epsilon \to 0$ to get
\begin{equation}
|v(0,t)| \le C ||\theta||_{L^\infty},
\end{equation}
where we have used the fact that~\cite{stein2}
\begin{equation}
\lim_{\epsilon \to 0}\frac{1}{2\epsilon}\int_{-\epsilon}^{\epsilon}|v(x,t)|=|v(0,t)|.
\end{equation}
In conclusion, we have shown that while the velocity remains bounded at the origin,
its first derivative goes to infinity in finite time, or what is the same, the fluid develops either a shock-wave or quasi-shock-wave. Finally, we can show that these singular solutions have finite kinetic energy. We know
that $||\theta||_{L^2}(t) \le ||\theta||_{L^2}(0)$, as proven in Ref.\cite{acordoba}. Using the Calderon-Zygmund inequality~\cite{majda} we can claim that
\begin{equation}
||v||_{L^2}(t) = ||H(\theta)||_{L^2}(t) \le \tilde{C} ||\theta||_{L^2}(t) \le \tilde{C} ||\theta||_{L^2}(0),
\end{equation}
for a finite positive constant $\tilde{C}$, and thus we see that the kinetic energy remains bounded for all times.

To finish our analysis of this model let us point out one very interesting feature of it. The particle
trajectories are defined, in this case, by
\begin{equation}
\frac{dx}{dt}=v=-H(\theta),
\end{equation}
what implies, for a particle starting at the origin, that
\begin{equation}
\frac{dx}{dt}=0,
\end{equation}
for all times, due to the antisymmetric character of $H(\theta)$. So one particle initially located at the
origin will stay there for all times. Since this particle is stopped, not only the velocity but also the
acceleration will vanish for all times, implying
\begin{equation}
\frac{d^2x}{dt^2}=\partial_t v+v\partial_xv=0.
\end{equation}
This is, the velocity obeys Burgers equation locally at the origin. This fact is very important, since it
shows that the pressure vanishes at the origin, and the shock is due to an artificial noninteracting
character of the particles. These considerations push us to conjecture that the shock-wave developed in this
type of flow is exactly the same as that formed in the Burgers equation.

\section{The hypoviscous Burgers equation}

While Eq.(\ref{ccf}) with $\mu=0$ develops (quasi-)shock-waves in finite time, it has been proven that adding a hypoviscosity to this equation results in global existence in time of the solution~\cite{acordoba}.
The corresponding hypoviscous equation reads~\cite{acordoba}
\begin{equation}
\partial_t \theta  - H(\theta) \partial_x \theta = \mu \Lambda^\alpha \theta,
\end{equation}
where $\mu>0$ is the viscosity, $1<\alpha<2$, and $\Lambda^\alpha$ is a fractional derivative of the Riesz type that is defined from its Fourier transform
\begin{equation}
(\Lambda^\alpha \theta)\hat{}=-|k|^\alpha\hat{\theta}.
\end{equation}
To confirm the analogue between this model and the Burgers equation we will analyze the hypoviscous version of the second
one, and we will prove global existence in time of the solutions. The effect of hypoviscosity on Burgers dynamics has been already studied in Ref.~\cite{bardos2}; however, in this reference it is analyzed the effect of the hypoviscous dissipativity on the Burgers Markovian Random Coupling Model, and we will analyze its effect directly on the Burgers equation.

The hypoviscous Burgers equation is
\begin{equation}
\label{hypo}
\partial_t v=-vv_x+\mu \Lambda^\alpha v.
\end{equation}
Obviously, the case $\alpha=2$ corresponds to the standard viscous Burgers equation.
In this case we will prove the nonexistence of the shock-wave or quasi-shock-wave for a situation similar to that of Eq.(\ref{ccf}), this is,
for two colliding fluid jets. We will suppose therefore that $v_x \le 0$ to simulate this scenario, and as assymptotic
conditions we will assume that $v_x$ is constant as $|x| \to \infty$. In the following proof,
the variables named as $C$, $C'$, and $\tilde{C}$ denote arbitrary constants and its value may change from
line to line.

Deriving once Eq.(\ref{hypo}) with respect to $x$ we find
\begin{equation}
\partial_t v_x =-(vv_x)_x+ \mu \Lambda^{\alpha}v_x.
\end{equation}
Now we want to calculate the $L^1$ norm of the velocity
\begin{equation}
\frac{d}{dt}||v_x||_{L^1}=-\int \partial_t v_xdx=\int(vv_x)_xdx+\mu \int \Lambda^{\alpha}v_xdx=0,
\end{equation}
where we have used our basic assumption $v_x \le 0$. So we have obtained the conservation in time of the $L^1$ norm
of $v_x$. With respect to the $L^2$ norm we have
\begin{eqnarray}
\nonumber
\frac{1}{2}\frac{d}{dt}||v_x||_{L^2}^2=\int v_x \partial_t v_x dx=-\int v_x(vv_x)_xdx+\mu\int v_x \Lambda^\alpha v_xdx= \\
=-\int v_x(vv_x)_xdx-\mu||\Lambda^{\alpha/2}v_x||_{L^2}^2.
\label{estvx}
\end{eqnarray}
Leibniz's rule, integration by parts and the application of the boundary conditions yield the two following equalities
\begin{equation}
-\int v_x(vv_x)_xdx=\int v_{xx}v_xvdx,
\end{equation}
\begin{equation}
-\int v_x(vv_x)_xdx=-\int v_x^3dx-\int v_{xx}v_xvdx,
\end{equation}
that can be combined to provide us with a new formulation of Eq.(\ref{estvx})
\begin{equation}
\frac{1}{2}\frac{d}{dt}||v_x||_{L^2}^2=\frac{1}{2} ||v_x||_{L^3}^3-\mu||\Lambda^{\alpha/2}v_x||_{L^2}^2.
\end{equation}
The third moment of $v_x$ might be estimated as follows
\begin{equation}
||v_x||_{L^3}^3 \le ||v_x||_{L^\infty}^2||v_x||_{L^1},
\end{equation}
now choose $\chi \in (1,\alpha)$, and use a Sobolev embedding to find
\begin{equation}
||v_x||_{L^\infty}^2 \le C(||v_x||_{L^2}^2+||\Lambda^{\chi/2} v_x||_{L^2}^2).
\end{equation}
We can now use the Fourier transform of $v_x$
\begin{equation}
(v_x)\hat{}(k)=\frac{1}{\sqrt{2\pi}}\int e^{ikx}v_x(x)dx
\end{equation}
to claim that
\begin{eqnarray}
\nonumber
||\Lambda^{\chi/2}v_x||_{L^2}^2=||(\Lambda^{\chi/2}v_x)\hat{}||_{L^2}^2=\int |k|^\chi |(v_x)\hat{}|^2 dk= \\
\nonumber
\int_{|k|\le R}|k|^\chi |(v_x)\hat{}|^2 dk+\int_{|k|\ge R}\frac{|k|^\alpha}{|k|^{\alpha-\chi}}|(v_x)\hat{}|^2 dk \le \\
R^{\chi}||v_x||_{L^2}^2+\frac{1}{R^{\alpha-\chi}}\int |k|^\alpha |(v_x)\hat{}|^2 dk=R^\chi||v_x||_{L^2}^2+
\frac{1}{R^{\alpha-\chi}}||\Lambda^{\alpha/2}v_x||_{L^2}^2,
\label{chile}
\end{eqnarray}
where we have used the isometry of the Fourier transform in $L^2$. We still need to estimate the second moment of $v_x$:
\begin{eqnarray}
\nonumber
||v_x||_{L^2}^2 \le ||v_x||_{L^1}||v_x||_{L^\infty}\le \frac{1}{2\epsilon}||v_x||_{L^1}^2+ \\
\frac{\epsilon}{2}||v_x||_{L^\infty}^2\le \frac{1}{2\epsilon}||v_x||_{L^1}^2+C\frac{\epsilon}{2}
\left(||v_x||_{L^2}^2+||\Lambda^{\alpha/2}v_x||_{L^2}^2\right),
\end{eqnarray}
where we have used a Sobolev embedding. Selecting $\epsilon$ small enough we are led to conclude
\begin{equation}
||v_x||_{L^2}^2\le \left(1-C\frac{\epsilon}{2}\right)^{-1}\left(\frac{1}{2\epsilon}||v_x||_{L^1}^2+
C'\frac{\epsilon}{2}||\Lambda^{\alpha/2}v_x||_{L^2}^2 \right).
\end{equation}
This inequality, in addition to Eq.(\ref{estvx}) yields
\begin{eqnarray}
\nonumber
||v_x||_{L^3}^3 \le C||v_x||_{L^1}\left[(1+R^\chi)\left(1-C'\frac{\epsilon}{2}\right)^{-1} \right. \times \\
\left(\frac{1}{2\epsilon}||v_x||_{L^1}^2+
\tilde{C}\frac{\epsilon}{2}||\Lambda^{\alpha/2}v_x||_{L^2}^2 \right)
+ \left. \frac{1}{R^{\alpha-\chi}}||\Lambda^{\alpha/2}v_x||_{L^2}^2\right],
\end{eqnarray}
and employing Eq.(\ref{chile}), choosing a sufficiently large $R$ and a
sufficiently small $\epsilon$, we arrive at the desired estimate
\begin{equation}
||v_x||_{L^2}\le C.
\end{equation}

Applying a second spatial derivative over Eq.(\ref{hypo}) we obtain
\begin{equation}
\partial_tv_{xx}=-(vv_x)_{xx}+ \mu \Lambda^\alpha v_{xx}.
\end{equation}
We can now compute the $L^2$ norm of $v_{xx}$
\begin{eqnarray}
\nonumber
\frac{1}{2}\frac{d}{dt}||v_{xx}||_{L^2}^2=-\int v_{xx}(vv_x)_{xx}dx+ \mu \int v_{xx}\Lambda^\alpha v_{xx}dx= \\
-\int v_{xx}(vv_x)_{xx}dx-\mu||\Lambda^{\alpha/2}v_{xx}||_{L^2}^2,
\end{eqnarray}
and by reiteratively using Leibniz's rule, integration by parts, and the boundary conditions we find that this equation
reduces to
\begin{equation}
\frac{1}{2}\frac{d}{dt}||v_{xx}||_{L^2}^2=-\frac{5}{2}\int v_x v_{xx}^2dx-\mu||\Lambda^{\alpha/2}v_{xx}||_{L^2}^2.
\end{equation}
The first integral in the right hand side of this equation may be estimated as follows
\begin{eqnarray}
\nonumber
\int v_{xx}^2v_x dx\le ||v_{xx}||_{L^\infty}||v_{xx}||_{L^2}||v_x||_{L^2} \le \\
\frac{\epsilon}{2}||v_{xx}||_{L^\infty}^2+\frac{1}{2\epsilon}||v_{xx}||_{L^2}^2||v_x||_{L^2}^2,
\end{eqnarray}
and we might continue this chain of inequalities by means of the Sobolev embedding
\begin{equation}
||v_{xx}||_{L^\infty}^2 \le C(||v_{xx}||_{L^2}^2+||\Lambda^{\alpha/2}v_{xx}||_{L^2}^2),
\end{equation}
to conclude
\begin{equation}
\int v_{xx}^2v_x dx\le \frac{\epsilon}{2}C(||v_{xx}||_{L^2}^2+||\Lambda^{\alpha/2}v_{xx}||_{L^2}^2)+
\frac{1}{2\epsilon}||v_{xx}||_{L^2}^2||v_x||_{L^2}^2.
\end{equation}
Choosing $\epsilon$ small enough and substituting this result in Eq.(\ref{hypo}) we find
\begin{equation}
\frac{d}{dt}||v_{xx}||_{L^2}^2 \le C\left(\frac{\epsilon}{2}+\frac{1}{2\epsilon}||v_x||_{L^2}^2 \right)||v_{xx}||_{L^2}^2,
\end{equation}
which yields
\begin{equation}
||v_{xx}||_{L^2}^2 \le Ce^{\tilde{C}t}.
\end{equation}
We can finish using this result in addition to the Sobolev inequality
\begin{equation}
||v_x||_{L^\infty}^2\le C(||v_x||_{L^2}^2+||v_{xx}||_{L^2}^2),
\end{equation}
to get the desired estimate
\begin{equation}
||v_x||_{L^\infty}\le C(1+e^{\tilde{C}t}),
\end{equation}
which prohibits the formation of shock-waves in finite time for the hypoviscous Burgers equation.

\section{News from two- and three-dimensional incompressible flows}

In order to understand the physical meaning of the possible blow-ups appearing in the Euler equations
it is useful to take a look to two- and three-dimensional incompressible flow. This is very important
to delucidate whether or not the one-dimensional models have physical meaning.

The problem of the blow-up for an incompressible fluid was studied in Ref.~\cite{cf}). They studied
four different equations: the Quasi-Geostrophic equation, the ideal two-dimensional Magneto-Hydrodynamics
equation, the two-dimensional Euler equation, and the Boussinesq equation. They found that for all these
equations, two arcs moving with the fluid cannot collapse in finite time into one single
arc, provided the velocity is bounded. This is a strong
result against the possible formation of (quasi-)shock-waves within this type of flow. Even in three-dimensional
incompressible ideal flow there is a similar result proven in this same direction~\cite{dcordoba}.

As we have seen, a (quasi-)shock-wave develops when two fluid particles collide in finite time. Two arcs moving with
the fluid are level sets of fluid particles, its mutual collapse would imply the formation of a
(quasi-)shock-wave. The nonexistence of this type of singularity for bounded velocities suggests that the
possible blow-up of the solution to the three-dimensional incompressible Euler equations is related to
the breakdown of the nonrelativistic description of the fluid, while it has no relation with the phenomenon
of turbulence. Furthermore, this fact indicates that the appearance of (quasi-)shock-waves in one-dimensional
models is more related to a mathematical artifact than to a real physical phenomenon.

\section{Conclusions}

In this work we have shown that one-dimensional models for three-dimensional incompressible hydrodynamics showing the
appearance of singular behaviour fail to reproduce some of the most important features of the flow. In the best cases,
the appearance of a (quasi-)shock-wave is due to the strong compressible character of the flow; actually, we arrive at the
absurd conclusion that these models for incompressible hydrodynamics are more compressible than a compressible flow. This is inferred from the fact that the shock develops due to the noninteracting character of
the fluid particles in its neighbourhood, what produces collisions among them.
This strongly suggests that one-dimensional settings are too limiting to correctly represent incompressible hydrodynamics.

On the other hand, the results involving two-dimensional incompressible flow show that the appearance of singularities is
related with a divergence of the velocity, what would rule out the possibility of (quasi-)shock-wave formation within the flow.
This would show that possible singularities have nothing to say in the transition to turbulence in incompressible flows.
Furthermore, the divergence of the velocity would indicate the breakdown of the nonrelativistic description of the fluid,
but would say nothing about the small scale properties of it. This indicates that a blow-up of the velocity would
have relevance in the study of some special fluids, such as interstellar plasmas, which reach velocities that may be
comparable to that of the light. However, the daily observed phenomenon of turbulence, which appears in common fluids at
velocities not comparable with light speed, seems to be not related with the existence of singular solutions to the
three-dimensional incompressible Euler equations.

\section*{Acknowledgments}

The author acknowledges Uriel Frisch for providing him with many useful references.
This work has been partially supported by the Ministerio de Educaci\'on y Ciencia (Spain) through Projects No. EX2005-0976 and FIS2005-01729.

\end{document}